\documentclass[10pt,aps,prb,showpacs,superscriptaddress,twocolumn,citeautoscript]{revtex4-1}

\usepackage{graphicx}  % needed for figures
\usepackage{dcolumn}   % needed for some tables
\usepackage{bm}        % for math 
\usepackage{amssymb}  
\usepackage{amsmath}
\usepackage{color,braket}

\newcommand{\beq}{\begin{equation}}
\newcommand{\eeq}{\end{equation}}

\newcommand{\fr}{\bm{r}}

\newcommand{\fK}{\bm{k}}
\newcommand{\fu}{\bm{u}}
\newcommand{\fv}{\bm{v}}

\newcommand{\fE}{\bm{E}}

\newcommand{\fH}{\bm{H}}
\newcommand{\fB}{\bm{B}}
\newcommand{\fD}{\bm{D}}
\newcommand{\fJ}{\bm{J}}

\newcommand{\eps}{\varepsilon}
\newcommand{\subeqs}[1]{\begin{align} #1 \end{align}}
\begin{document}
\title{Strong Spatial Dispersion in Time-Modulated Dielectric Media }
\author{Dani Torrent}
\email{dtorrent@uji.es}
\affiliation{GROC, UJI, Institut de Noves Tecnologies de la Imatge (INIT), Universitat Jaume I, 12071, Castell\'o, (Spain)}
\date{\today}
%%%%%%%%%%%%%%%%%%%%%%%%%%%%%%%%%%%%%%%%%%%%%%%%%%%%%%%%%%%%%%%%%%%%%%%%%%%%%%%%%%%%%%%%%%%

%%%%%%%%%%%%%%%%%%%%%%%%%%%%%%%%%%%%%%%%%%%%%%%%%%%%%%%%%%%%%%%%%%%%%%%%%%%%%%%%%%%%%%%%%%%
\begin{abstract}
We present an effective medium description of time-modulated dielectric media. By taking the averaged fields over one modulation period, the relationship between them is derived, defining therefore the different constitutive parameters. In the most general situation, it is found that the effective material is described by means of a spatially and temporally dispersive transverse dielectric function and a constant longitudinal dielectric function. It has been also found that the frequency dependence in the former is weak, in comparison with its wavenumber-dependence (spatial dispersion). Different physical consequences of this spatial dispersion are discussed, with special emphasis in the weak dispersion approximation, limit in which it is found that the effective material behaves as a resonant and isotropic magnetodielectric medium with no additional longitudinal mode, as it is commonly found in spatially dispersive materials. Time-dependent media opens therefore an alternative way of designing dynamically tunable metamaterials.

\end{abstract}
%%%%%%%%%%%%%%%%%%%%%%%%%%%%%%%%%%%%%%%%%%%%%%%%%%%%%%%%%%%%%%%%%%%%%%%%%%%%%%%%%%%%%%%%%%%
\maketitle
%%%%%%%%%%%%%%%%%%%%%%%%%%%%%%%%%%%%%%%%%%%%%%%%%%%%%%%%%%%%%%%%%%%%%%%%%%%%%%%%%%%%%%%%%%%M
\section{Introduction}
%%%%%%%%%%%%%%%%%%%%%%%%%%%%%%%%%%%%%%%%%%%%%%%%%%%%%%%%%%%%%%%%%%%%%%%%%%%%%%%%%%%%%%%%%%%

The study of naturally or artificially structured materials is a classic problem in physics and engineering. Also named composites, a countless number of theoretical and experimental methods have been developed to properly understand their properties\cite{milton2003theory}. In this context, metamaterials are a special type of composites, where the effective medium has extreme constitutive parameters mainly due to local resonances of the constitutive elements\cite{simovski2007local,alu2011first}. For electromagnetic materials, these extreme parameters implies the existence of simply or double negative materials, where a huge literature exist concerning their physics and applications\cite{cui2010metamaterials,cai2010optical,marques2011metamaterials}. However, the existence of strongly resonant effective properties is usually accompanied of more or less weakly spatially dispersive properties\cite{cuabuz2008spatial,kruk2012spatial,yaghjian2013homogenization,chern2013spatial,torrent2015resonant}.

Metamaterials have recently evolved towards more complex structures, and the possibility of design new materials based on the temporal-modulation of the constitutive parameters has also been explored. Then, it can be shown that, when a given medium presents  time-modulated constitutive parameters, some effects like non-reciprocity, gain and tunability can be easily achieved\cite{hayrapetyan2016electromagnetic,nassar2017non,torrent2018loss,torrent2018nonreciprocal,chen2019nonreciprocal,trainiti2019time,croenne2019non}.

The concept of effective medium for a time-modulated material is similar to that of a spatially-modulated\cite{pacheco2020effective,huidobro2020homogenisation}, in the sense that, when the modulation frequency (spatial or temporal) is fast and the operating wavelength and frequency cannot detect that modulation, we detect an effective medium with some averaged constitutive parameters. When the parameters are spatially modulated, we obtain an effective medium which a strong temporal dispersion (frequency-dependence) but a weak spatial dispersion (wavenumber-dependence). 

In this work it will be shown that the temporal modulation exchanges these properties, and the effective medium has a strong spatial dispersion and a weak temporal dispersion. It will be shown that, after averaging the electromagnetic fields, the temporal modulation of the dielectric constant results in an effective material with a non-local transverse dielectric function but a local longitudinal one. Analytical expressions will be derived for these two functions and some examples will be analyzed. Finally, it will be shown how in the weak dispersion approximation the effective material behaves as an isotropic magnetodielectric medium, a property achieved so far mainly by complex three-dimensional metamaterials\cite{simovski2009model,muhlig2011self,ponsinet2015resonant,gomez2016hierarchical}. Therefore, the temporal modulation of the dielectric constant is an excellent alternative for the realization of complex materials with extreme electromagnetic properties.

The paper is organized as follows: After this introduction, Sec. \ref{sec:eps} presents the homogenization method and the expressions for the non-local dielectric function. Section \ref{sec:epsk} analyzes the spatially dispersive dielectric function and some of its properties. Then, in Sec. \ref{sec:ST} the consequences of the spatial dispersion for finite slabs in both space and time are discussed and in Sec. \ref{sec:mu} the artificial magnetic effect due to the weak dispersion approximation is analyzed. Finally, Sec. \ref{sec:sum} summarizes the work.

%%%%%%%%%%%%%%%%%%%%%%%%%%%%%%%%%%%%%%%%%%%%%%%%%%%%%%%%%%%%%%%%%%%%%%%%%%%%%%%%%%%%%%%%%%%M
\section{Non-local dielectric function}
\label{sec:eps}
%%%%%%%%%%%%%%%%%%%%%%%%%%%%%%%%%%%%%%%%%%%%%%%%%%%%%%%%%%%%%%%%%%%%%%%%%%%%%%%%%%%%%%%%%%%

The evolution of the electromagnetic field in matter excited by an external current  $\fJ_{ext}$ and charge density $\rho_{ext}$ is described by means of Maxwell's equations,
\subeqs{
\nabla\times\fE&=-\partial_t\fB, \label{eq:rotE}\\
\nabla\times\fH&=\partial_t\fD+\fJ_{ext}, \label{eq:rotB} \\
\nabla\cdot \fD&=\rho_{ext}\label{eq:divD}\\
\nabla\cdot\fB&=0
}
whose solution can be obtained only once we know the constitutive equations relating the fields $\fH,\fB,\fE$ and $\fD$. In the media we aim to study, these constitutive equations are the corresponding ones to a non-magnetic material with a time dependent electrical permittivity $\eps(t)$, then we will have that $\fE(\fr,t)=\eps^{-1}(t)\fD(\fr,t)$ and $\fB(\fr,t)=\mu_0\fH(\fr,t)$. 

 We will assume as well that the function $\epsilon(t)$ is $T$-periodic in time, and that the modulation frequency $\nu_M=1/T$ is larger than the operating frequency $\omega$ of the external currents and charges. We can assume then that the response of the system will be a smooth function modulated by a fast function whose average in a period $T$ will be zero. The relationship between these averaged fields will define the effective constitutive parameters of the material, identically as it happens in spatially periodic media. 

Then, let us assume that the time-dependent dielectric function $\eps(t)$ and its inverse $\eps^{-1}(t)$ can be expanded in a Fourier series of the form
\beq
f(t)=\sum_nf_ne^{-2i\pi nt/T}
\eeq
where $f_n$ are labeled $\eps_n$ and $\eps_{n}^{-1}$ for $\eps(t)$ and $\eps^{-1}(t)$, respectively. Notice that with this notation for $n=0$ we have that $\eps_0=\braket{\eps(t)}$ and $\eps_0^{-1}=\braket{\eps(t)^{-1}}$, not to be confused with the permittivity of vacuum. 

The external current and density are the ones selecting the operating frequency $\omega$ and wavenumber $\fK$, thus we assume
 \subeqs{
 \fJ_{ext}&=\fJ_0e^{i\fK\cdot\fr}e^{-i\omega t},\label{eq:Jext}\\
\rho_{ext}&=\rho_0e^{i\fK\cdot\fr}e^{-i\omega t},
}
and we know that in this case the solution for the fields will be of the form
\begin{multline}
\label{eq:ut}
\fu(\fr,t)=e^{i\fK\cdot\fr}e^{-i\omega t}\sum_n\fu_ne^{-2i\pi nt/T}=\\e^{i\fK\cdot\fr}e^{-i\omega t}\fu_0+e^{i\fK\cdot\fr}e^{-i\omega t}\sum_{n\neq0}\fu_ne^{-2i\pi nt/T},
\end{multline}
where $\fu=\fE,\fD,\fB$ and $\fH$. Therefore, the response of the electromagnetic field to an external field of wavenumber $\fK$ and frequency $\omega$ is composed of a slow component $\fu_0$ and a fast modulation $\fu_n$ for $n\neq 0$. For a fast modulation frequency $\nu_M=1/T$ we can interpret the $n=0$ as the averaged ``observable'' terms, so that their evolution will define the evolution of the effective material.

Then, the relationship between the $n=0$ terms in the above expansions is
\subeqs{
\fK\times\fE_0&=\omega\fB_0, \label{eq:rotE0}\\
\fK\times\fH_0&=-\omega\fD_0-i\fJ_{0}, \label{eq:rotH0} \\
i\fK\cdot \fD_0&=\rho_{0},\\
\fK\cdot\fB_0&=0.
}
These expressions show that the $n=0$ component of the field expansion satisfies Maxwell's equation in Fourier space, as expected. The objective now is to find the relationship between these components, which will define 
the effective constitutive parameters of the medium. The effective magnetic permeability can be trivially found, since
\beq
\fB_0=\mu_0\fH_0,
\eeq
although it will be seen later that, due to the spatial dispersion in the effective dielectric constant, an effective magnetic permeability will be found. Concerning the relationship between $\fE_0$ and $\fD_0$, we see that this can be written as
\beq
\label{eq:E0Dm}
\fE_0=\eps(t)^{-1}=\eps_0^{-1}\fD_0+\sum_{n\neq 0}\eps_{-n}^{-1}\fD_n,
\eeq
thus we need the relationship between $\fD_n$ and $\fD_0$ in order to properly define an effective constitutive equation. This relationship is found from Maxwell's equations, since the wave equation for $\fD$ is
\beq
\eps^{-1}(t)\nabla\times\nabla\times\fD=-\mu_0\partial^2_{tt}\fD-\mu_0\partial_t\fJ_{ext},
\eeq
which, after using equations \eqref{eq:Jext} and \eqref{eq:ut}, is equivalent to
\beq
-\sum_{m}\eps_{n-m}^{-1}\fK\times\fK\times\fD_m=\mu_0\Omega_n^2\fD_n+i\mu_0\omega\fJ_0\delta_{n0}
\eeq
where we have defined the displaced frequency $\Omega_n=\omega+\frac{2n\pi}{T}$. For $n=0$ the above equation is 
\beq
\label{eq:D0J0}
-\eps_0^{-1}\fK\times\fK\times\fD_0+\sum_{m\neq 0}\eps_{-m}^{-1}k^2\fD_m=\mu_0\omega^2\fD_0+i\mu_0\omega\fJ_0
\eeq
since we have from equation \eqref{eq:divD} that $i\fK\cdot\fD_n=\delta_{n0}\rho_0$. Similarly, for $n\neq0$ we have
\beq
-\eps_n^{-1}\fK\times\fK\times\fD_0+\sum_{m\neq 0}\eps_{n-m}^{-1}k^2\fD_m=\mu_0\Omega^2_n\fD_n,
\eeq
from which we obtain the desired relationship between $\fD_m$ and $\fD_0$,
\beq
\label{eq:Dm}
\fD_m=-\sum_{n\neq0}\Gamma_{mn}^D\eps_n^{-1}\fK\times\fK\times\fD_0,
\eeq
with
\beq
\label{eq:GD}
\Gamma_{mn}^D=\left(\mu_0\Omega_n^2\delta_{mn}-\eps^{-1}_{n-m}k^2\right)^{-1}.
\eeq
We have obtained therefore the fundamental relationship between the fast terms and the average field $\fD_0$, thus equation \eqref{eq:Dm} can now be introduced into \eqref{eq:E0Dm} and we get, after some little algebra, 
\beq
\label{eq:ED}
\fE_0=\eps^{-1}_T(\omega,\fK)\fD_0+(\eps_L^{-1}-\eps^{-1}_T(\omega,\fK))\fu_{\fK}\fD_0\cdot\fu_{\fK}
\eeq
where we have defined the transverse and longitudinal inverse dielectric constants $\eps_T^{-1}$ and $\eps_L^{-1}$, respectively, as
\subeqs{
\eps^{-1}_T(\omega,\fK)&=\braket{\eps^{-1}}+k^2\sum_{n,m\neq0}\eps_{-m}^{-1}\Gamma_{mn}^D\eps_{n}^{-1},\label{eq:em1}\\
\eps^{-1}_L&=\braket{\eps^{-1}},
}
which clearly satisfies
\subeqs{
\fK\times\fE_0&=\eps_T^{-1}(\omega,\fK)\fK\times\fD_0\label{eq:eT}\\
\fK\cdot\fE_0&=\eps_L^{-1}\fK\cdot\fD_0.
}
We see therefore that the effective material behaves as a material with a non-local transverse dielectric response but a local one. The local response is independent of both the frequency and the wavenumber, while the transverse dielectric function depends on both of them, i.e., it is non-local in space and time. However, as can be seen from equation \eqref{eq:GD}, this dependence is of the form $\omega+2n\pi/T$, for $n\neq0$, which actually implies that if the modulation frequency $\nu_M=1/T$  of the medium is larger than the operating frequency, this dependence will disappear and we will have non-locality in space only. This is the opposite situation as that tipically found in classical periodic materials, where non-locality in time (frequency-dependent dielectric constant) is stronger than non-locality in space, for identical reasons as those considered here, except for layered or wire media, as will be discussed later.

We can check the consistency of the definition of $\eps_T^{-1}$ by inserting equation \eqref{eq:Dm} into equation \eqref{eq:D0J0}, since we obtain
\beq
\label{eq:rotrotD}
-\eps^{-1}_T(\omega,\fK)\fK\times\fK\times\fD_0=\mu_0\omega^2\fD_0+i\mu_0\omega\fJ_0,
\eeq
which is identical to the wave equation we would obtain from equations \eqref{eq:rotE0} and \eqref{eq:rotH0} using the constitutive equation derived in \eqref{eq:ED}.

We can proceed in a similar way to obtain an alternative expression for $\eps_T(\omega,\fK)$. Thus, the wave equation for the $\fH$ field is given by
\beq
\label{eq:wave1}
\nabla\times\nabla\times\fH=-\mu_0\partial_t(\epsilon(t)\partial_t\fH)+\nabla\times \fJ_{ext},
\eeq
again using equations \eqref{eq:Jext} and \eqref{eq:ut} it becomes
\beq
k^2\fH_n=\mu_0\sum_m\Omega_n\eps_{n-m}\Omega_m\fH_m+i\fK\times\fJ_0\delta_{n0}.
\eeq
As before, we can split these equations into the $n=0$ component,
\beq
\label{eq:H0J0}
k^2\fH_0=\mu_0\braket{\eps}\omega^2\fH_0+\mu_0\omega \sum_{m\neq 0} \eps_{-m}\Omega_m\fH_m+i\fK\times\fJ_0
\eeq
and the $n\neq0$ set of equations,
\beq
k^2\fH_n=\mu_0\omega\eps_{n}\Omega_n\fH_0+\mu_0\sum_{m\neq 0} \Omega_n\eps_{n-m}\Omega_m\fH_m,
\eeq
from which we can obtain the expression analogue of \eqref{eq:Dm} but for $\fH_m$,
\beq
\fH_m=\mu_0\omega\sum_{n\neq0}\Gamma^H_{mn}\Omega_n\eps_{n}\fH_0,
\eeq
where now we have defined,
\beq
\label{eq:GH}
\Gamma^H_{mn}=\left(k^2\delta_{mn}- \mu_0\Omega_n\eps_{n-m}\Omega_m\right)^{-1}.
\eeq
Equation \eqref{eq:H0J0} is now
\beq
k^2\fH_0=\mu_0\eps_T(\omega,\fK)\omega^2\fH_0+i\fK\times\fJ_0,
\eeq
where
\beq
\label{eq:e}
\eps_T(\omega,\fK)=\braket{\eps}+\mu_0\sum_{m,n\neq0}\Omega_n\eps_n\Gamma^H_{nm}\eps_{-m}\Omega_m.
\eeq
We arrive therefore to two possible definitions of $\eps_T(\omega,\fK)$, as given from equations \eqref{eq:em1} and \eqref{eq:e}, and we would like to check if these are equivalent or not, and to do so we should prove that $\eps_T(\omega,\fK)\eps_T(\omega,\fK)^{-1}=1$. Then, if we use the time-dependent constitutive equation and introduce it in equation \eqref{eq:rotB} we get
\beq
\nabla\times\fD=-\mu_0\eps(t)\partial_t\fH,
\eeq
which in $\omega-\fK$ space is
\beq
\fK\times\fD_0=\omega\mu_0\eps_{0}\fH_0+\mu_0\sum_{m\neq0}\Omega_m\eps_{-m}\fH_m=\omega\mu_0\eps_T(\omega,\fK)\fH_0.
\eeq
However, from equations \eqref{eq:eT} and \eqref{eq:rotE0} we get
\beq
\eps_T^{-1}(\omega,\fK)\fK\times\fD_0=\omega\mu_0\fH_0,
\eeq
which actually implies that
\beq
\eps^{-1}_T(\omega,\fK)\eps_T(\omega,\fK)=1,
\eeq 
as expected. 

We can now use the expressions of equation \eqref{eq:em1} or \eqref{eq:e} for our better convenience, as will be shown below. For instance, if we want to determine the limit of low $\omega$ and $\fK$, what we could call the ``static limit'', it is more suitable to use \eqref{eq:em1}, so that we obtain trivially
\beq
\eps_T^{-1}=\braket{\eps^{-1}}=\eps_L^{-1},
\eeq
the material is then a homogeneous dielectric material with an averaged reciprocal dielectric constant, with identical transverse and longitudinal responses. The above result generalizes the result obtained in \cite{pacheco2020effective} for a layered time-dependent material, showing therefore the consistency of this approach.

%%%%%%%%%%%%%%%%%%%%%%%%%%%%%%%%%%%%%%%%%%%%%%%%%%%%%%%%%%%%%%%%%%%%%%%%%%%%%%%%%%%%%%%%%%%M
\section{Frequency-independent spatially dispersive dielectric function}
\label{sec:epsk}
%%%%%%%%%%%%%%%%%%%%%%%%%%%%%%%%%%%%%%%%%%%%%%%%%%%%%%%%%%%%%%%%%%%%%%%%%%%%%%%%%%%%%%%%%%%
Let us consider now the case of a weak modulation changing the dielectric constant from $\braket{\eps}+\Delta$ to $\braket{\eps}-\Delta$ harmonically at a frequency $\nu_M=1/T$. We have therefore
\beq
\eps(t)=\braket{\eps}+\Delta\cos(2\pi\nu_Mt).
\eeq
We can assume first that $2\pi\nu_M>>\omega$, to focus the analysis on the spatially dispersive properties of the material. Since all the Fourier components for $n\neq0,\pm1$ are zero, we can restrict our analysis to these orders. It is clear now that the $\Gamma^H$ matrix is diagonal with elements
\beq
\Gamma_{\pm1\pm1}^H=\frac{1}{k^2-\mu_0\braket{\eps}4\pi^2\nu_M^2},
\eeq
therefore the effective dielectric constant is given by
\beq
\label{eq:eweak}
\eps_T(\fK)=\braket{\eps}+\frac{\Delta^2}{2\braket{\eps}}\frac{k_M^2}{k^2-k_M^2},
\eeq
where $k_M^2=4\pi^2\nu_M^2\mu_0\braket{\eps}$.

The above dielectric constant is similar to that found for the so-called ``wire media'', in which propagation takes place parallel to a periodic distribution of cylinders\cite{maslovski2002wire,belov2003strong,maslovski2009nonlocal}. This similarity has a clear explanation: in any periodic medium, we will have factors of the form $k+2\pi/a$ in the homogeneization process, so that the dependence on $k$ will be in general small. However, since in wire media the periodicity along the $z$ axis is ``broken'' by letting $a\to\infty$, we have a strong dependence on $k$. Similarly, the time periodicity of the material is generally ``broken'', for that reason we have a strong dependence on $\omega$ in composites. However, in this case the ``broken'' periodicity is along the full space, while we still have periodicity along the time axis, thus we find a strong $k$-dependence and, consequently, strong non-locality. 

The functional form of equation \eqref{eq:eweak} is not unique of the weak harmonic modulation, as will be demonstrated below. The matrix $\Gamma^H$ is defined in equation \eqref{eq:GH} as the inverse of the matrix $M$ given by
\beq
M_{nm}=k^2\delta_{mn}- \mu_0\Omega_n\eps_{n-m}\Omega_m.
\eeq
Since this matrix is Hermitian, we can use the eigendecomposition of a matrix and express the inverse as
%\beq
%\Gamma^H=\sum_\ell \frac{\fv_\ell \otimes \fv_\ell^\dag}{\lambda_\ell},
%\eeq
a function of the eigenvectors $\fv_\ell$ and eigenvalues $\lambda_\ell$ of $\bm{M}$. Since matrix $\bm{M}$ has the form $\bm{M}=\bm{I}k^2-\bm{\chi}$, its eigenvalues are $\lambda=k^2-\lambda'$, with $\lambda'$ being the k-independent eigenvalues of $\bm{\chi}$, thus we have that
\beq
\Gamma^H=\sum_\ell \frac{\fv_\ell \otimes \fv_\ell^\dag}{k^2-\lambda_\ell'},
\eeq
and equation \eqref{eq:e} is 
\beq
\label{eq:ekfull}
\eps_T(\fK)=\braket{\eps}+\mu_0\sum_\ell \sum_{m,n\neq0}\Omega_n\eps_n\frac{\fv_\ell \otimes \fv_\ell^\dag}{k^2-\lambda_\ell'}\eps_{-m}\Omega_m.
\eeq
In the denominator of the above expression the $\lambda_\ell'$ are independent of $k$, and it can be easily shown that the numerator is as well independent of $k$, since
\beq
\frac{\partial \fv_\ell}{\partial k}=\left(\lambda_\ell\bm{I}-\bm{M}\right)^\dag\frac{\partial \bm{M}}{\partial k}\fv_\ell=0.
\eeq
Then, equation \eqref{eq:ekfull} shows that the general form of $\eps(\fK)$ is similar to the weak harmonic modulation but with more poles. In the above expressions we have ignored the possible dependence on frequency of $\eps$, however we have previously discussed that this dependence is weak, but if it has to be included it will appear through the eigenvalues and eigenvectors of the expansion.

%%%%%%%%%%%%%%%%%%%%%%%%%%%%%%%%%%%%%%%%%%%%%%%%%%%%%%%%%%%%%%%%%%%%%%%%%%%%%%%%%%%%%%%%%%%M
\section{Space-time representation of finite materials}
\begin{figure}[h!]
	\centering
	\includegraphics[scale=0.9]{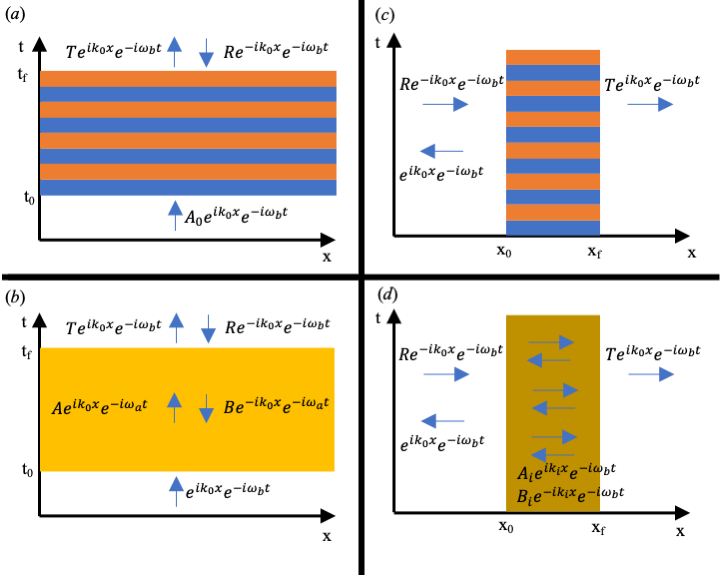}
	\caption{Space-time representation of different modulation scenarios. Panel $a$: Homogeneous material modulated in time during a period $t=t_f-t_0$. Panel $b$: Homogenization of the periodic modulation presented in panel $a$. Panel $c$: Periodic modulation in time of a finite slab placed between $x=x_0$ and $x=x_f$. Panel $d$: Homogenization of the structure presented in panel $d$.}
	\label{fig:figure1}
\end{figure}

%%%%%%%%%%%%%%%%%%%%%%%%%%%%%%%%%%%%%%%%%%%%%%%%%%%%%%%%%%%%%%%%%%%%%%%%%%%%%%%%%%%%%%%%%%%
\label{sec:ST}
It is worth now to discuss some physics concerning the time  modulation of the dielectric constant, which will help us to understand the possible implications of equation \eqref{eq:eweak}. Let us consider the situation illustrated in figure \ref{fig:figure1} panel $a$. What we see is a spatially homogeneous material with some dielectric constant $\eps_b$ and, at $t=t_0$, a periodic modulation is applied until $t=t_f$. This situation is similar to that analyzed for acoustic waves in [\onlinecite{hayrapetyan2013propagation}] and [\onlinecite{torrent2018loss}]. We can assume that we are faraway the band gap, where the material would be unstable, and that the conditions for the application of the effective medium condition holds, then we are in the situation described in panel $b$ of figure \ref{fig:figure1} . We can see how for $t<t_0$ a wave is propagating through the material with some frequency $\omega_b$ and wavenumber $k_b$. Once the modulation begins, we excite a ``transmitted'' and ``reflected'' wave, but the wavenumber of these waves continues being $k_b$, and it is the frequency the quantity that has changed\cite{torrent2018loss,pacheco2020effective}. To obtain the new frequency in the effective material we need to solve the dispersion relation 
\beq
\omega^2=\eps_T^{-1}(\fK)k^2,
\eeq
where we have assumed that there is no dependence of $\eps$ on $\omega$. Thus, the time-slab has a similar behaviour than the spatial-slab, since once the modulation stops we will have a transmitted and reflected waves whose relative amplitude can present gain or loss, as explained in [\onlinecite{torrent2018loss}], but spatial dispersion does not changes the physics of the problem, the only thing that changes is its resonant-like nature, similarly as for spatially modulated materials.

%%%%%%%%%%%%%%%%%%%%%%%%%%%%%%%%%%%%%%%%%%%%%%%%%%%%%%%%%%%%%%%%%%%%%%%%%%%%%%%%%%%%%%%%%%%%%%%%%%%%%%%%
%%%%%%%%%%%%%%%%%%%%%%%%%%%%%%%%%%%%%%%%%%%%%%%%%%%%%%%%%%%%%%%%%%%%%%%%%%%%%%%%%%%%%%%%%%%%%%%%%%%%%%%%

However, the common situation is to have a spatially limited material (slab), even if we have an additional modulation in time. Let us consider therefore the situation shown in figure \ref{fig:figure1}, panel $c$, where we see a periodically modulated material in time, but limited to the region $x\in[x_0,x_f]$. The full wave analysis of this situation is rather complex, as can be seen for the geometry of the domains involved, however it is now where the effective medium concept is specially useful, as it is shown in panel $d$ of figure \ref{fig:figure1}. The problem now is limited to a classical transmission-reflection problem, but now the operating frequency $\omega_b$ is indeed a conserved quantity, since any time dependence of the geometry has been averaged. An additional difficulty appears in this case, since spatial dispersion usually involves additional propagating modes that requires the use of additional boundary conditions, as studied in many works\cite{ting1975electrodynamics,halevi1982additional,silveirinha2009additional,maslovski2010generalized}, although a recent approach based on an elastodynamic model for spatially dispersive materials could be more adequate for isotropic strongly dispersive materials with a transverse dielectric response\cite{alvarez2020generalized}.

%%%%%%%%%%%%%%%%%%%%%%%%%%%%%%%%%%%%%%%%%%%%%%%%%%%%%%%%%%%%%%%%%%%%%%%%%%%%%%%%%%%%%%%%%%%M
\section{Isotropic artificial magnetism}
%%%%%%%%%%%%%%%%%%%%%%%%%%%%%%%%%%%%%%%%%%%%%%%%%%%%%%%%%%%%%%%%%%%%%%%%%%%%%%%%%%%%%%%%%%%
\label{sec:mu}

The only treatable situation in which additional boundary conditions are not required is the so called ``weak'' dispersion approximation, which assumes that $k$ is a small quantity so that we can expand $\eps_T(\fK)$ as
\beq
\eps_T(\fK)\approx\eps_T(0)+\gamma k^2,
\eeq
since in our case it is clear that there is no linear term in $k$. Then, according to our response model, we have
\beq
\fD_0=\eps_T(\fK)+(\eps_L-\eps_T(\fK))\fu_{\fK}\fD_0\cdot\fu_{\fK},
\eeq
we showed before that for $\fK=\bm{0}$ and $\omega=0$ we have $\eps_T=\eps_L$, so that the above expression is approximated to
\beq
\fD_0\approx\eps_T(\bm{0})\fE_0-\gamma \fK\times\fK\times\fE=\eps_T(\bm{0})\fE_0-\omega\gamma \fK\times\fB_0,
\eeq
which gives an ``artificial'' isotropic magentic response $\mu(\omega)$
\beq
\mu(\omega)=\frac{\mu_0}{1-\mu_0\gamma\omega^2}.
\eeq
It is worth to mention that isotropic artificial magnetism has been a topic of intense research in the domain of metamaterials, specially at optical frequencies, and complex three-dimensional structures are in general required to achieve this interesting property\cite{simovski2009model,muhlig2011self,ponsinet2015resonant,gomez2016hierarchical}. The temporal modulation of the dielectric constant is therefore an interesting alternative, although it presents different and obvious technical difficulties. 

Taylor expanding \eqref{eq:eweak} it is easy to see that, for a weak periodic modulation,
%\beq
%\gamma=-\frac{\Delta^2}{\braket{\eps}k_M^2},
%\eeq
%and some little algebra shows
\beq
\mu(\omega)=\frac{\mu_0}{1-\frac{\Delta^2}{\braket{\eps}^2}\frac{\omega^2}{\omega_M^2}}.
\eeq

Finally, equation \eqref{eq:em1} allows to obtain a very nice expression of the Taylor expansion of $\eps_T^{-1}(\fK)$ in the general case, since it is easy to show
\beq
\eps_T^{-1}(\fK)\approx\braket{\eps^{-1}}+k^2\sum_{n\neq0}\frac{|\eps_n^{-1}|^2}{\mu_0\Omega_n^2},
\eeq
from which we obtain the expression for $\gamma$ in the most general case,
\beq
\gamma=-\frac{1}{\braket{\eps^{-1}}^2}\sum_{n\neq0}\frac{|\eps_n^{-1}|^2}{\mu_0\Omega_n^2}.
\eeq

%%%%%%%%%%%%%%%%%%%%%%%%%%%%%%%%%%%%%%%%%%%%%%%%%%%%%%%%%%%%%%%%%%%%%%%%%%%%%%%%%%%%%%%%%%%M
\section{Summary}
%%%%%%%%%%%%%%%%%%%%%%%%%%%%%%%%%%%%%%%%%%%%%%%%%%%%%%%%%%%%%%%%%%%%%%%%%%%%%%%%%%%%%%%%%%%
\label{sec:sum}

In summary, we have derived an effective medium theory for time-modulated dielectric materials. It has been found that, in general, the fields can be decomposed into an averaged and fast modulated components, and the relationship between the slow components of the fields define the effective parameters of the material. It has been then demonstrated that the effective dielectric constant has a transverse component presenting a strong spatial dispersion but a weak temporal one, contrarily as space-modulated metamaterials, where the dominant effect is temporal dispersion. Analytical expressions have been derived for several examples, and the consequences of this strong spatial dispersion have been discussed under different scenarios, with especial emphasis in the so-called ``weak dispersion approximation'', in which an artificial isotropic magnetic response has been found. Since the modulation frequency is, in principle, a dynamic quantity easier to control in real time than the spatial modulation, we consider that this approach opens the door to a new class of dynamically tunable metamaterials.

%Let us consider now that the solution for the propagation of waves in such effective material, i.e., when no sources are present. This can be obtained from the condition
%\beq
%k^2=\omega^2\mu_0\eps(\fK)
%\eeq
%if we define the averaged wavenumber $k_0^2=\mu_0\braket{\eps}\omega^2$, using equation \eqref{eq:eweak} the above equation is
%\beq
%k^4-(k_M^2+k_0^2)k^2+\left(1-\frac{\Delta^2}{2\braket{\eps}^2}\right)k_0^2k_M^2=0
%\eeq
%which has four roots for a given frequency $\omega$, corresponding to two pairs of backwards and forwards propagating modes. These solutions are
%\beq
%k^2=\frac{k_M^2+k_0^2}{2}\pm\sqrt{\frac{(k_M^2-k_0^2)^2}{4}-\frac{\Delta^2}{2\braket{\eps}}k_0^2k_M^2}.
%\eeq
%Further algebra on the above equation shows that, assuming 
%\beq
%\eeq

%%%%%%%%%%%%%%%%%%%%%%%%%%%%%%%%%%%%%%%%%%%%%%%%%%%%%%%%%%%%%%%%%%%%%%%%%%%%%%%%%%%%%%%%%%%
%%%%%%%%%%%%%%%%%%%%%%%%%%%%%%%%%%%%%%%%%%%%%%%%%%%%%%%%%%%%%%%%%%%%%%%%%%%%%%%%%%%%%%%%%%%
\begin{acknowledgments}
Daniel Torrent acknowledges financial support through the ``Ram\'on y Cajal'' fellowship under grant number RYC-2016-21188 and to the Ministry of Science, Innovation and Universities through Project No. RTI2018- 093921-A-C42. 
\end{acknowledgments}
%\bibliographystyle{apsrev}
%\bibliography{bibliography}


\end{document}